\documentclass[fleqn]{article}
\usepackage[a4paper,margin=25.4mm]{geometry}
\usepackage[english]{babel}

\usepackage{amsmath}
\usepackage{amssymb}
\usepackage{sistyle}
\usepackage{cite}

\usepackage{tabularx}

\usepackage{graphicx}			
\usepackage{wrapfig}			
\usepackage[
	labelfont={sf,bf},			
	textfont={sf},			
	font=small,					
	justification=RaggedRight	
	]{caption}
\usepackage{subcaption}			
\usepackage{url}
\usepackage{xcolor}

\usepackage[unicode=true,pdfusetitle, bookmarks=false, breaklinks=false,pdfborder={0 0 0},backref=false,colorlinks=false] {hyperref}

\begin{document}

\begin{center}
	\hfill\\[1.0cm]
	{\Large\textbf{Alignment conditions of the human eye for few-photon vision experiments}}\\[0.2cm]
	{\large{T.H.A. van der Reep$\left.^{1,*}\right.$, and
	W. L\"offler$\left.^{1}\right.$}} \\[0.2cm]
	{$\left.^{1}\right.$\emph{Leiden Institute of Physics, Niels Bohrweg $2$, $2333$ CA Leiden, The Netherlands}}\\[0.1cm]
	$\left.^{*}\right.$reep@physics.leidenuniv.nl\\[0.2cm]
	\today
\end{center}

\begin{abstract}
In experiments probing human vision at the few-photon level, precise alignment of the eye is necessary such that stimuli reach the most sensitive region of the retina. However, in literature there seems to be no consensus on the optimal eye alignment for such experiments. Typically, experiments are performed by presenting stimuli nasally or temporally, but the angle under which the few-photon pulses are presented varies between $7^{\circ}$ and $23^{\circ}$. Here we combine a $3$-dimensional eye model with retinal rod density measurements from literature in a ray tracing simulation to study the optimal eye alignment conditions and necessary alignment precision. We find that stimuli, directed at the eye's nodal point, may be best presented under an inferior angle of $12.6^{\circ}$ with respect to the visual axis. Defining a target area on the retina with a radius of $\SI{0.5}{mm}$ around the optimum location, we find the horizontal and vertical angular precision should be better than $0.90^{\circ}$ given a horizontal and vertical translational precision of $\pm\SI{1}{mm}$ and a depth translational precision of $\pm\SI{5}{mm}$.
\end{abstract}

\section{Introduction}\label{secIntroduction}
Since the advancement of quantum mechanics human vision is studied at the few-photon level. In the $1940$s it was shown for the first time that the human visual system is sensitive to light pulses containing only a few photons \cite{Hechtetal1941}. In later experiments this limit has been reproduced for narrow \cite{Velden1946,Sakitt1972,Teichetal1982} and wide \cite{DentonPirenne1954,Deyetal2021} field stimuli, and recently it has been shown that humans might even perceive single photons \cite{Tinsleyetal2016}. This result has been challenged by an experiment supporting the hypothesis that at least two photons are necessary to invoke visual perception \cite{Kilpelainenetal2024}.  On the other hand, flux measurements indicate a threshold for vision of $\SI{0.1}{photons/ms}$ \cite{MarriottMorrisPirenne1959}. These results are quite remarkable, given the optical loss in the human eye, the dark activation rate of human photoreceptor cells, and the noisy environment of the brain. To probe the few-photon response of the human visual system further, we recently proposed to perform quantum detector tomography on the human visual system in the few-photon range \cite{Reepetal2023}.\\
Experiments probing the absolute limits of human vision use stimuli that target the areas of the retina where rod cells are abundant. Rod cells are used to mediate scotopic (low-light) vision and are known to be sensitive to single photon stimulation \cite{Phanetal2014}. Most experiments target the nasal or temporal retinal area. The only exception in the works cited above is Ref. \cite{Kilpelainenetal2024}, in which experiments were performed in the vertical visual field. However, there seems to be no consensus on the angle under which light stimuli should presented to the test subjects -- chosen values of this angle range from $7^{\circ}$ to $23^{\circ}$. With exception from Ref. \cite{Tinsleyetal2016} stating that the density of rod cells is highest at their chosen angle of $23^{\circ}$ in the temporal visual field (nasal retina), typically no rationale is given for choosing a specific angle. Although it has previously been shown that the threshold for vision at the temporal retina is lowest for angles between approximately $10^{\circ}$ and $15^{\circ}$ \cite{HalettMarriottRodger1962}, it should be noted that the cited works do not provide an in-depth discussion of eye and stimulus alignment. To the best of our knowledge, such a discussion is not presented in literature to date, which makes an actual comparison between the different angle values hard.\\
In this work, we will discuss eye alignment for few-photon vision experiments from a theoretical point of view. Using ray tracing simulations in a $3$-dimensional model of the human eye, we show that the $10^{\circ}$ to $15^{\circ}$ range of minimum threshold coincides with the high-density rod (hDR) region in the temporal retina \cite{Curcioetal1990}. For this reason we propose to target the retinal highest-density rod (HDR) region, which is found in the superior area of the retina \cite{Curcioetal1990,JonasSchneiderNaumann1992}. Using our ray tracer we can determine the stimulus conditions for targeting the HDR region and determine the necessary alignment precision for such experiments. We will proceed by reviewing the geometry of the human eye in section \ref{secEyeGeom}. Then, in section \ref{secSimulation} we will discuss the developed simulation model, after which the results are presented in section \ref{secResults}. We conclude by a discussion and conclusion in sections \ref{secDiscussion} and \ref{secConclusions}, respectively.

\section{Human eye geometry}\label{secEyeGeom}
\begin{figure}
\centering
	\includegraphics[width=\textwidth]{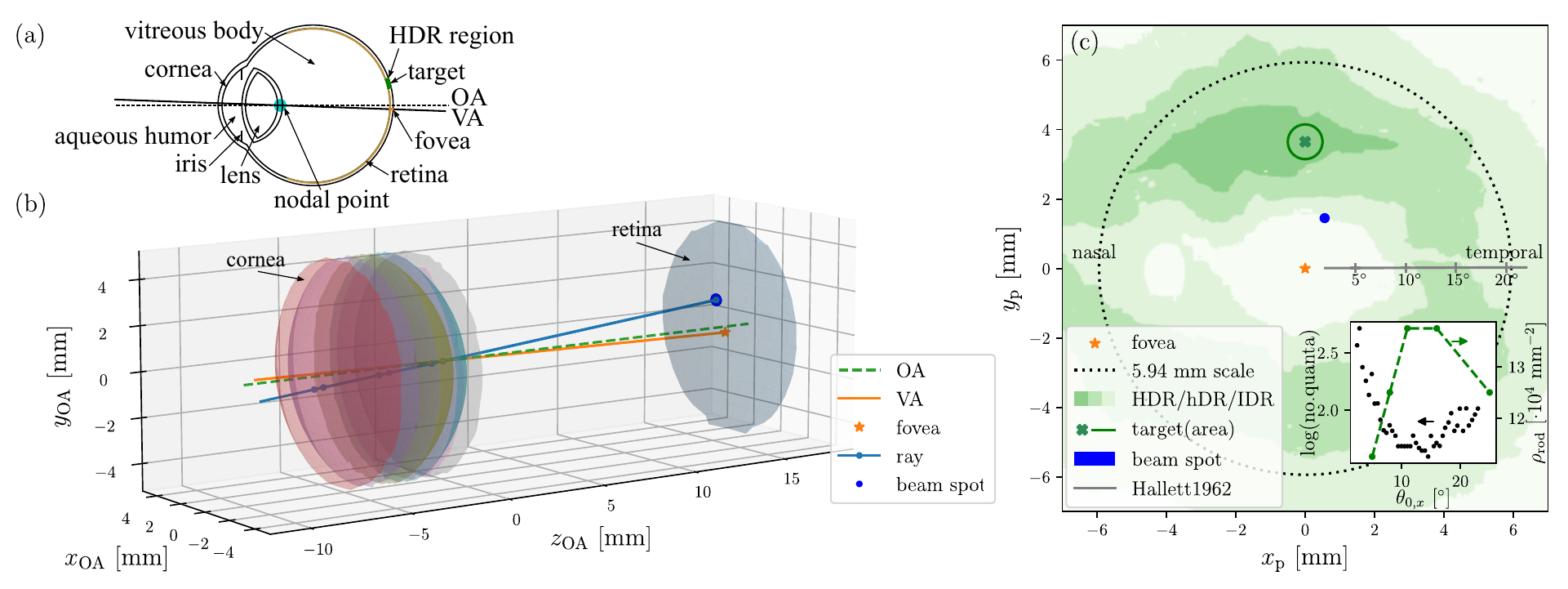}
	\caption{The human eye and simulation model. (a) Schematic overview (sagittal [side] view) of the human eye according to Gullstrand's exact eye model with relevant parts indicated (see text). (b) Beam transmission $\vec{\theta}_0=[2,5]^{\circ}$, $\Delta r_{\text{m}}=\vec{0}$, $\theta_{1/2}=0.5^{\circ}$) through the $3$-dimensional Gullstrand model. The spherical surfaces represent the various refractive elements in the human eye and the central ray as well as the spot of the beam are shown, highlighting the ray-surface interaction points. At each intersection, Snell's law is applied to obtain the new ray direction. (c) The beam spot on the retina in perimetric view, along with the HDR region, hDR ring and IDR region (extracted from Ref. \cite{Curcioetal1990}), and the target area (see text). Due to the $\SI{5.94}{mm}$-scale (black dotted circle), positions are directly comparable with the data from Ref. \cite{Curcioetal1990}. The gray line shows the experimental region of Ref. \cite{HalettMarriottRodger1962} according to our simulation. The figure inset shows the visual threshold (in number of photons) as a function of stimulus eccentricity (left axis, extracted from Ref. \cite{HalettMarriottRodger1962}), as compared to rod density (right axis). The dashed line is intended to guide the eye.}\label{figEyeModel}
\end{figure}

The human eye is an intricate optical instrument enabling vision. In this section we shortly review the eye's geometry, and the regions and points of interest for eye alignment and vision experiments. We base our discussion on Gullstrand's exact eye model \cite{Gullstrand1909}, depicted in figure \ref{figEyeModel}(a), to which we add an iris/pupil and a visual axis. The iris is modelled as a $\SI{1000}{mm}$-radius spherical surface directly in front of the eye lens. The refractive parameters of this model are given in table \ref{tabGullstrand}. Although the Gullstrand model is one of the earliest models, developed in $1909$, we consider it a good balance between accuracy and complexity.

\begin{table}[b]
	\centering
	\caption{Model parameters of the Gullstrand eye model with added iris. $R_{\text{ant (post)}}$ is the anterior (posterior) element radius. $z_{\text{ant}}$ is the position of the element's anterior as measured from the cornea along the OA. $t$ is the thickness and $n$ the index of refraction.}\label{tabGullstrand}
	\begin{tabularx}{0.74\textwidth}{|l|c|c|c|c|X|}
	\hline
		Element& $R_{\text{ant}}$ [mm]& $R_{\text{post}}$ [mm] &$z_{\text{ant}}$ [mm]&$t$ [mm]&$n$ [-]\\
	\hline\hline
		cornea & $7.7$&$6.8$ & $0$ &$0.50$ &$1.376$\\
		aqueous humour & $6.8$ &$10.0$ &$0.50$ &$3.10$ &$1.336$\\
		(iris) & $1000$ & NA & $3.60$ & NA & NA\\
		ant. lens cortex & $10.0$ &$7.911$ & $3.60$ &$0.546$ &$1.386$\\
		lens core &$7.911$ &$-5.76$ &$4.146$ &$2.419$ &$1.406$\\
		post. lens cortex & $-5.76$ &$-6.0$ & $6.565$ &$0.635$ &$1.386$\\
		vitreous body & $-6.0$ &$-12.0$ &$7.20$ &$16.80$ &$1.336$\\
		retina &$-12.0$ & NA & $24.0$ & NA &NA\\
 \hline\hline
	\end{tabularx}
\end{table}

Upon entering the eye, light travels through the cornea, aqueous humour, eye lens and vitreous body before reaching the retina. In the Gullstrand model, the eye lens is divided in three parts, thus capturing its non-constant refractive index \cite{Khanetal2018}. For the pupil we assume a diameter of $\SI{8}{mm}$, the pupil diameter of dark-adapted eyes \cite{Lazaretal2024}.\\
The retina contains photo-receptive cells: cones and rods. Cones mitigate photopic (colour) vision and are primarily found in the fovea (yellow spot) central in the visual field \cite{Curcioetal1990}. On the other hand, rods are utilised in scotopic (low light) vision and are absent in the fovea. Instead, the rod density peaks in a ring around the fovea, the hDR ring ($1.25\times 10^5$ to $\SI{1.375e5}{rods/mm^2}$), and is found to be highest in the retinal section superior to (above) the fovea, the HDR region ($1.375\times 10^5$ to $\SI{1.625e5}{rods/mm^2}$) \cite{Curcioetal1990}. Specifically, the HDR region is located $2.93$ to $\SI{4.34}{mm}$ superior to the fovea measured along the vertical retinal meridian. It has a width of approximately $\SI{5}{mm}$. In this work we will target the point of this region $\SI{3.65}{mm}$ superior to the fovea and we define a target area around this point with a radius of $\SI{0.5}{mm}$. Although the target and target area are not located in the middle of the HDR region, we consider the superior section of the retina for easier experimental implementation.\\
In the following, two ocular axes are important: the optical axis (OA) and the visual axis (VA). The OA is the eye's axis of symmetry. Interestingly, this axis does not co-align with the VA of the human eye. The latter axis connects the fovea with the object one is viewing. With respect to the OA, the VA is generally misaligned by an angle of $\alpha_x=5^{\circ}$ to $6^{\circ}$ temporally (towards the temple) and $\alpha_y=2^{\circ}$ to $3^{\circ}$ inferiorly (downward, see figure \ref{figEyeModel}(a)) \cite{Tscherning1920}, although $\alpha_y$ can found to be $0^{\circ}$ or directed superiorly as well \cite{Tscherning1920,Kimetal2018}. The OA and VA cross in the eye's nodal point \cite{Simpson2022}, see figure \ref{figEyeModel}(a). The nodal point is defined such that light rays entering the eye under an angle $\vec{\theta}_0=[\theta_{0,x},\theta_{0,y}]$ leave the eye lens under the same angle, and can therefore be thought of as the eye's optical centre.

\section{Eye model simulation}\label{secSimulation}
To model the transmission of light stimuli through the eye, we implement Gullstrand's model in a $3$-dimensional eye ray tracing simulation of the right eye in \textsc{Python}. Within the simulation, the different refractive surfaces are modelled as spherical surfaces refracting incoming rays by application of Snell's law, see figure \ref{figEyeModel}(b). It should be noted that, using this approach, any eye model based on spherical surfaces can be straightforwardly implemented.\\
In our simulation, we will use a right-handed reference system fixed to the nodal point, which we find to be located $\SI{7.1}{mm}$ posterior of the cornea. The $z$-direction is aligned with the VA, the temporal direction is the positive $x$-direction and the superior direction is the positive $y$-direction. The direction $\vec{\theta}=[\theta_{x},\theta_y]$ of a light ray is taken to be positive for rays propagating in positive $z$-, $x$- and $y$-direction. Here, $\theta_x$ lies in the $xz$-plane and $\theta_y$ in the $yz$-plane. Within this reference system and for the results presented below, we set $\vec{\alpha}=[\alpha_x,\alpha_y]=[5,-2]^{\circ}$, in accordance with the values stated in section \ref{secEyeGeom}.\\

The input of our model is the initial ray direction $\vec{\theta}_0$ (i.e., the source orientation with respect to the eye) and the light source position misalignment $\Delta\vec{r}_{0}=[\Delta x_{0},\Delta y_{0},\Delta z_{0}]$ in VA coordinates. This defines the parameter space $(\vec{\theta}_0,\Delta\vec{r}_0)$ to be explored. For $\Delta\vec{r}_{0}=\vec{0}$ the ray is directed at the nodal point and for $\vec{\theta}_0=\Delta\vec{r}_{0}=\vec{0}$ the ray is co-aligned with the VA reaching the fovea. We transform the VA input coordinates to OA coordinates and perform ray tracing in two steps. First, the intersection point of the ray and the next optical surface is calculated. Second, Snell's law is applied at this intersection point to obtain the new ray direction. By consecutively performing these calculations for all optical surfaces in the model, we find the location at which the ray intersects the retina. These coordinates are back-transformed from OA coordinates to VA coordinates to yield the retinal impact position $\vec{r}_{\text{ret}}=[x_{\text{ret}},y_{\text{ret}},z_{\text{ret}}]$. If the ray misses a surface, hits the iris or shows internal reflection according to Snell's law, it is removed from further consideration. Iris hits are detected, whenever the radial distance of the ray-iris interaction point to the OA exceeds the iris radius $R_{\text{ir}}=\SI{4}{mm}$.\\
Retinal distances between two retina locations $\vec{r}_{\text{ret},1}$ and $\vec{r}_{\text{ret},2}$, $d(\vec{r}_{\text{ret},1},\vec{r}_{\text{ret},2})$, are calculated using the great-circle distance over the retinal sphere. This distance is evaluated to determine whether the ray hits or misses the target area [by considering $d(\vec{r}_{\text{ret}},\vec{r}_{\text{tar}})$], and to present the retinal impact position in a perimetric view of the retina as employed in Ref. \cite{Curcioetal1990}. In this view, see figure \ref{figEyeModel}(c), the radial distance from the fovea is non-distorted, whereas distortion takes place in azimuthal direction. The perimetric coordinates are given by $[x_{\text{p}},y_{\text{p}}]=d(\vec{r}_{\text{ret}},\vec{r}_{\text{fov}})[\cos(\phi),\sin(\phi)]$, where $\phi$ is the azimuthal angle as measured counter-clockwise from the positive $x_{\text{p}}$-axis, such that $\tan(\phi)=y_{\text{ret}}/x_{\text{ret}}$.

\section{Results}\label{secResults}
In this section the results of our simulations are presented, while we gradually proceed in parameter space complexity: from a simulation of a single ray and beam we will arrive at considering the full parameter space $(\vec{\theta}_0,\Delta\vec{r}_0)$ via the sub-parameter spaces $\theta_{0,x}$, $\theta_{0,y}$, $\vec{\theta}_0$ and $\Delta\vec{r}_0$.\\ 

First we calculate the retinal beam spot size of a light beam with a half-opening angle of $\theta_{1/2}=0.5^{\circ}$, offset by $\vec{\theta}_0=[2,5]^{\circ}$ from the VA and focussed on the nodal point ($\Delta\vec{r}_{0}=\vec{0}$), thus representing the Maxwellian view \cite{Westheimer1966} commonly applied in few-photon vision experiments. We note that ray optics is sufficient to calculate the beam spot size at the retina, which follows from a consideration of the beam's Rayleigh range. This is the range around the focal point where the wave properties of light cannot be neglected. For a beam with a Gaussian profile, a wavelength $\lambda=\SI{500}{nm}$ (at which the rods' detection efficiency is highest) and a refractive index of $n=1.35$, the Rayleigh range equals $z_{\text{R}}=\lambda/(\pi\theta_{1/2}^2 n)=\SI{1.5}{mm}$. Since we consider larger distances here, wave properties can be neglected.\\
In figure \ref{figEyeModel}(b) the central beam ray is highlighted ($\vec{\theta}_0=[2,5]^{\circ}$) showing the interaction points with the refractive surfaces of the eye. Figure \ref{figEyeModel}(c) shows a perimetric view of the retina including the central fovea, hDR ring, the HDR region, the intermediate density rod (IDR) region ($1.125\times 10^5$ to $\SI{1.25e5}{rods/mm^2}$) \cite{Curcioetal1990}, the target area and the $\SI{5.94}{mm}$-scale from Ref. \cite{Curcioetal1990}. As can be seen, the simulated beam spot has a radius of $\SI{0.15}{mm}$ on the retina. We found that the spot size is constant with input angle $\vec{\theta}_0$ (not shown), implying a full spot can be simulated by the central beam ray only.\\

To support our proposal to target the superior HDR region, we compare the rod density measurements from Ref. \cite{Curcioetal1990} to visual threshold data from Ref. \cite{HalettMarriottRodger1962}. We vary $\theta_{0,x}$ ($\theta_{0,y}=0$, $\Delta\vec{r}_0=\vec{0}$) to obtain the retinal impact locations for light stimuli entering the eye under angles of $2^{\circ}$ to $23^{\circ}$ as used in Ref. \cite{HalettMarriottRodger1962}. This experimental region has been indicated in figure \ref{figEyeModel}(c) as a gray line. The inset of this figure shows the visual threshold as well as the rod density in this region, indicating a strong negative correlation between these two parameters.\\

\begin{figure}
\centering
	\begin{minipage}[t]{0.495\textwidth}
	\includegraphics[width=\textwidth]{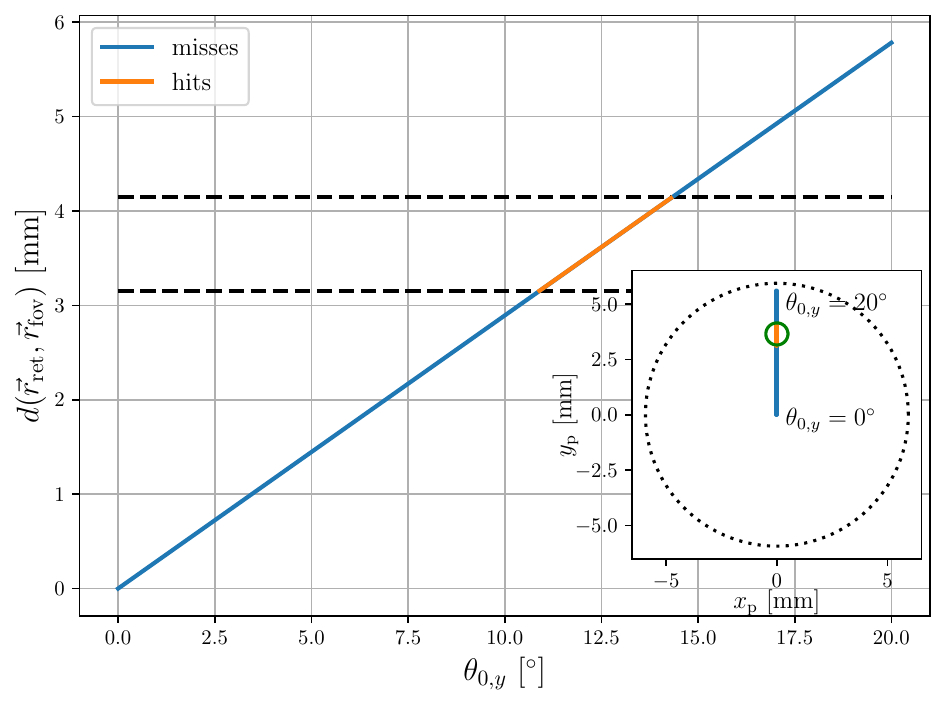}
	\caption{Determination of the optimal input angle $\theta_{0,y}$ ($\theta_{0,x}=0$, $\Delta\vec{r}_0=\vec{0}$). The inset shows the retinal impact position of rays in perimetric view, while varying $\theta_{0,y}$ from $0^{\circ}$ to $20^{\circ}$, featuring hits and misses. Calculating the great-circle distance between the fovea and the ray's retinal impact location, we find the optimum angle equals $12.6^{\circ}$ and the target area is hit for angles in between $10.9^{\circ}$ and $14.3^{\circ}$, as indicated by the dashed lines.}\label{figAngley}
	\end{minipage}\hfill
	\begin{minipage}[t]{0.495\textwidth}
	\includegraphics[width=\textwidth]{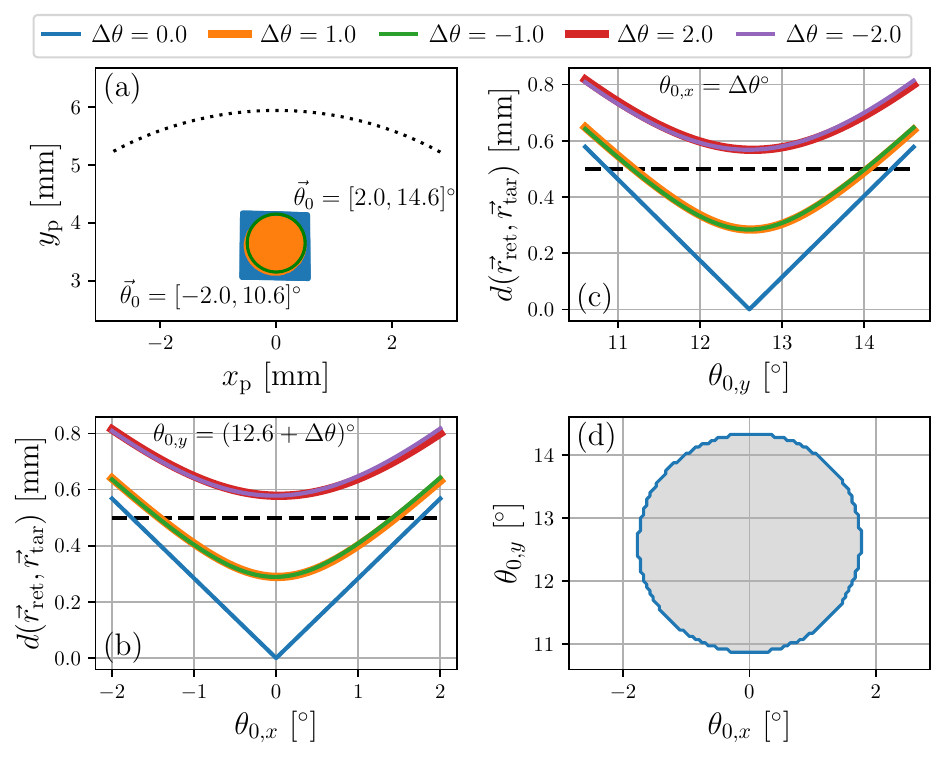}
	\caption{Necessary angular precision for $\Delta\vec{r}_0=\vec{0}$. (a) Hits and misses varying $\theta_{0,x}$ from $-2^{\circ}$ to $2^{\circ}$ and $\theta_{0,y}$ from $10.6^{\circ}$ to $14.6^{\circ}$ in perimetric view. (b) and (c) show the great-circle distance between the retinal impact location and the target as a function of $\theta_{0,x}$ ($\theta_{0,y}$ constant, b) and $\theta_{0,y}$ ($\theta_{0,x}$ constant, c). The target area is indicated using a dashed line. (d) Parameter space plot of $\vec{\theta}_0$. Rays with $\vec{\theta}_0$ originating from the shaded region hit the target area.}\label{figAngularPrec}
	\end{minipage}
\end{figure}

By varying $\theta_{0,y}$ ($\theta_{0,x}=0$, $\Delta \vec{r}_{0}=\vec{0}$) we can determine the optimum input angle for hitting the target. As can be seen in figure \ref{figAngley}, increasing $\theta_{0,y}$ translates the retinal impact position in superior direction, as expected. The optimum angle to hit the target equals $13.1^{\circ}$. $\theta_{0,y}$ may deviate by $1.7^{\circ}$ from this value, such that the ray hits the target area.\\

In figure \ref{figAngularPrec} the full angular parameter space $\vec{\theta}_0$ is considered ($\Delta\vec{r}_0=\vec{0}$). Figure \ref{figAngularPrec}(a) depicts the hitting and missing rays in perimetric view. Since $\Delta{\vec{r}}=\vec{0}$, no lens distortion is observed. Figures \ref{figAngularPrec}(b) and (c) show the retinal distance between the ray's retinal impact point and the target as a function of $\theta_{0,x}$ and $\theta_{0,y}$ for several fixed values of $\theta_{0,y}$ and $\theta_{0,x}$, respectively. From these data, we depict the parameter space plot of $\vec{\theta}_0$ in figure \ref{figAngularPrec}(d). Rays originating from within the plotted boundary (the shaded area) hit the target area, whereas rays from outside the boundary miss it.\\

\begin{figure}
\centering
\includegraphics[width=\textwidth]{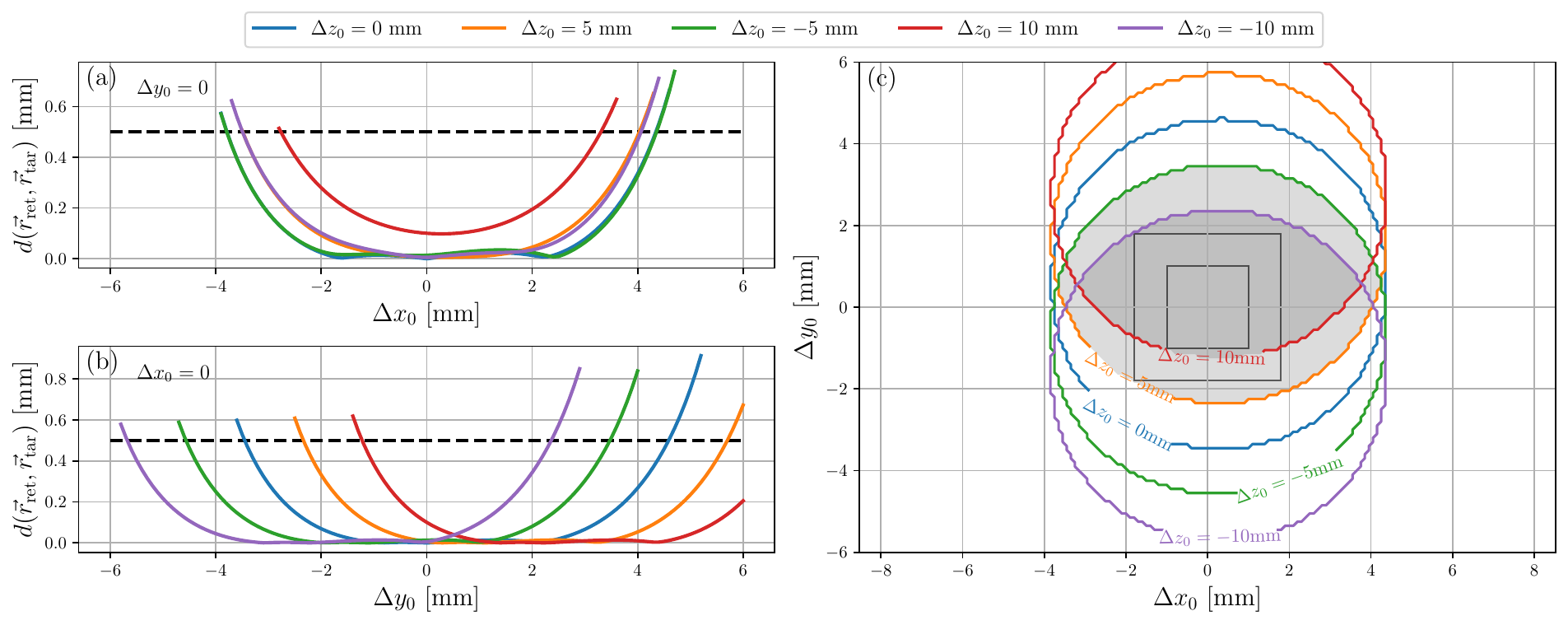}
\caption{Necessary translational precision for $\vec{\theta}_0=[0,12.6]^{\circ}$. (a) and (b) show the great-circle distance between the retinal impact location of the ray and the target for several values of $\Delta z_0$ as a function of $\Delta x_{0}$ ($\Delta y_0=0$, a) and $\Delta y_{0}$ ($\Delta x_0=0$, b). The target area is indicated by the dashed line. (c) depicts the parameter space plot of $\Delta\vec{r}_0$. If the precision in $\Delta z_0=\pm\SI{10}{mm}$, rays originating  from the region shaded darkgray hit the target area. If one would consider equal precision in $\Delta x_0$ and $\Delta y_0$, the region within the small square remains yielding a precision of $\pm\SI{1.0}{mm}$ for both $\Delta x_0$ and $\Delta y_0$. In case the precision in $\Delta z_0=\pm\SI{5}{mm}$, the lightgray region represents the allowed parameter space and the large square yields the equal $\Delta x_0$, $\Delta y_0$ precision to be $\pm\SI{1.8}{mm}$.}\label{figTransPrec}
\end{figure}

The same analysis can be performed considering only the translational parameter space $\Delta\vec{r}_0$ ($\vec{\theta}_0=[0,12.6]^{\circ}$). The results of this analysis are shown in figure \ref{figTransPrec}. In figures \ref{figTransPrec}(a) and (b) the retinal distance from the ray's impact position to the target can be observed for several values of $\Delta z_0$, given $\Delta y_0=0$ and $\Delta x_0=0$, respectively. As can be seen in these figures, for low values of $\Delta z_0$, a horizontal misalignment leads only to a small deviation from the target, see figure \ref{figTransPrec}(a). Only for larger misalignment this deviation increases substantially. The same holds for a vertical misalignment, see figure \ref{figTransPrec}(b). However, here it is also observed that misalignment in $y$- and $z$-direction are coupled. This is a direct result from the source orientation of $\vec{\theta}_0=[0,12.6]^{\circ}$. Figure \ref{figTransPrec}(c) depicts the full parameter space plot of $\Delta\vec{r}_0$. For the different values of $\Delta z_0$, rays originating from the parameter space $(\Delta x_0,\Delta y_0)$ inside the corresponding contour hit the target area. The same coupling as mentioned previously, between $\Delta y_0$ and $\Delta z_0$, is visible. In order to interpret this result further, let us assume that the alignment precision in $z$-direction is $\pm\SI{10}{mm}$, corresponding to $\Delta z_0=\pm\SI{10}{mm}$. In this case, rays from the parameter space spanned by both the contours $\Delta z_0=\SI{10}{mm}$ and $\Delta z_0=-\SI{10}{mm}$, shaded dark gray in figure \ref{figTransPrec}(c), hit the target area. If we further require the precision in $x$- and $y$-direction to be equal and centred around $0$, the allowed parameter space shrinks to the small square. This gives a required precision in $x$- and $y$-direction of $\pm\SI{1.0}{mm}$, given the precision in $z$-direction is $\pm\SI{10}{mm}$. Performing the same analysis for a $z$-precision of $\pm\SI{5}{mm}$ leads to a minimum required precision of $\SI{1.8}{mm}$ in $x$- and $y$-direction, as represented by the large square in figure \ref{figTransPrec}(d).\\

\begin{figure}
\centering
\includegraphics[width=0.5\textwidth]{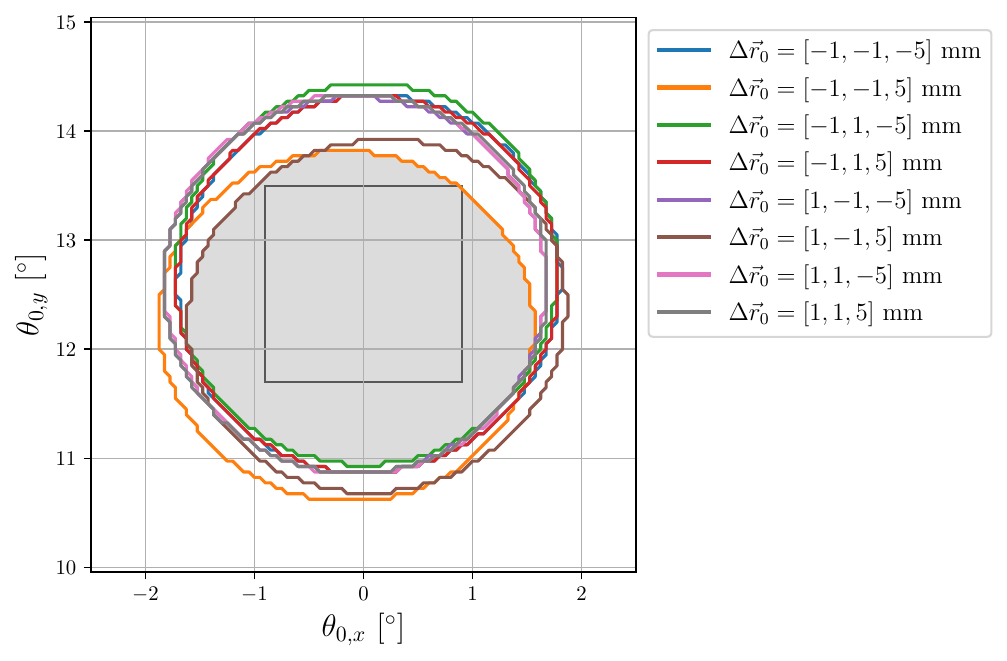}
\caption{Full $(\vec{\theta}_0,\Delta\vec{r}_0)$-parameter space precision. Assuming the precision in $x$- and $y$-direction is $\pm\SI{1}{mm}$, and the precision in $z$-direction equals $\pm\SI{5}{mm}$, the allowed parameter space of $\vec{\theta}_0$ is represented by the shaded area. Requiring the precision of $\theta_{0,x}$ and $\theta_{0,y}$ to be equal and centred around $[0,12.6]^{\circ}$, the remaining allowed $\vec{\theta}_0$ parameter space is indicated using a square. This leads to a necessary angular precision of $0.90^{\circ}$ in both $\theta_{0,x}$ and $\theta_{0,y}$.}\label{figFullPrec}
\end{figure}

As a last step, we combine the angular and translational parameter space to study the necessary alignment precision in the full $(\vec{\theta}_0,\Delta\vec{r}_0)$-parameter space. Assuming a precision in $x$- and $y$-direction of $\pm\SI{1}{mm}$, and in $z$-direction of $\pm\SI{5}{mm}$, the resulting angular parameter space from which rays hit the target area is presented in figure \ref{figFullPrec}. This figure shows the eight contours corresponding to $\Delta\vec{r}_0=[\pm 1,\pm 1,\pm 5]\SI{}{mm}$ in $\vec{\theta}_0$ space, and, as before, the parameter space for which these contours overlap yields the necessary precision for $\vec{\theta}_0$. This precision is indicated by the shaded area in figure \ref{figFullPrec}. Requiring the precision in $\theta_{0,x}$ and $\theta_{0,y}$ to be equal and centred around $\vec{\theta}_0=[0,12.6]^{\circ}$, the allowed parameter space is indicated by the square. This gives a necessary angular precision of $\theta_{0,x}$ and $\theta_{0,y}$ of $0.90^{\circ}$.

\section{Discussion}\label{secDiscussion}
Gullstrand's exact eye model captures many of the anatomical features of the human eye and is based on spherical refractive surfaces. To this model we added an iris and misalignment angle $\vec{\alpha}$ between OA and VA to resemble actual eyes further. However, in reality the eye's refractive surfaces are not quite spherical, and it is also known that the eye lens is not co-aligned with the OA, but rather makes a small angle with this axis, generally referred to as the angle $\kappa$ \cite{Kimetal2018}. Apart from this, the refractive index of the eye lens shows smoother variation than captured by Gullstrand's model \cite{Khanetal2018}. However, although the Gullstrand model is chosen out of a plethora of different eye models, we expect our results to be hardly dependent on eye model, since we base our analysis on deviations with respect to rays crossing the nodal point.\\ 

A slight influence of the VA-OA misalignment angle $\vec{\alpha}$ on the allowed parameter space is found. This angle is the cause that the shaded areas in figures \ref{figTransPrec}(c) and \ref{figFullPrec}, representing the allowed parameter space for hitting the target area, are not symmetrical about $[\Delta x_0,\Delta y_0]=[0,0]$ and $\vec{\theta}_0=[0,12.6]^{\circ}$, respectively. Repeating the analysis in figure \ref{figFullPrec} for $\vec{\alpha}=[6,-3]^{\circ}$ (the largest common misalignment, according to \cite{Tscherning1920}), we find that the required precision in $\theta_{0,x}$ and $\theta_{0,y}$ is still $0.90^{\circ}$, whereas for $\vec{\alpha}=[5.5,3.1]^{\circ}$ as determined for Korean eyes \cite{Kimetal2018}, the required angular precision equals $0.75^{\circ}$.\\

Comparison of visual threshold and rod density measurements suggests that targeting the HDR region is beneficial for stimulus detection probability in few-photon experiments. This region is found in the superior area of the retina. With our model, the optimum $\vec{\theta}_0$ is found to be $[0,12.6]^{\circ}$. This angle deviates by the angle commonly chosen in literature for low-photon vision experiments in which most often the nasal and/or temporal section are targeted.  Although in these sections hDR regions are present, the density of rods is largest in the HDR region. Measurements similar to those of Ref. \cite{HalettMarriottRodger1962} could be performed for the inferior-superior meridian of the retina. Thus it can be found whether using the superior part of the retina indeed shows a lower vision threshold than the temporal-nasal meridian, as we expect.\\
 
As noted, $\Delta\vec{r}_0$ represents the light source's position misalignment and during experiments $\vec{\theta}_0$ of the source is fixed. This implies that a test whether the combination of optical apparatus and participant fulfills the required precision to hit the HDR target area breaks down into three questions. First, it should be studied, how well the optical apparatus can be aligned to the participant with fully stabilised head and second, the stability of the head fixation in dark conditions must be studied. This gives the experimental $\Delta\vec{r}_0$, as used for obtaining the results in figure \ref{figFullPrec}. The third step in this process is to study how well the participant can fixate his/her gaze, e.g., using a dim fixation light commonly used in few-photon vision experiments. However, the values we find, a required angular precision of $0.85^{\circ}$ given a translational precision of $\SI{1}{mm}$ in $x$- and $y$-direction, and a $\SI{5}{mm}$ precision in $z$-direction do seem very feasible values \cite{kallmark2008}.\\

\section{Conclusions}\label{secConclusions}
We have studied the alignment conditions for few-photon vision experiments. To this end we implemented Gullstrand's exact eye model \cite{Gullstrand1909} with added iris and misalignment between the optical and visual axis in $3$ dimensions, and combined this model with rod density measurements of the retina \cite{Curcioetal1990} and visual threshold measurements \cite{HalettMarriottRodger1962}. Using our model we found a strong negative correlation between visual threshold and rod density, suggesting few-photon experiments should target the highest-density rod region. In order to target this region, we direct the beam at the eye's nodal point under the optimum angle of $0^{\circ}$ in the horizontal plane and $12.6^{\circ}$ in the inferior sagittal (from below in the vertical) plane with respect to the visual axis. Apart from this we studied the necessary alignment precision, for which we defined a target area with a $\SI{0.5}{mm}$ radius around the target point $\SI{3.65}{mm}$ superior to the fovea. From our model, it follows that the required angular precision equals $0.90^{\circ}$ for a translational precision of $\SI{1}{mm}$ in $x$-and $y$-direction and $\SI{5}{mm}$ in $z$-direction.

\subsection*{Acknowledgements}
We acknowledge funding from NWO (NWA.$1418$.$24$.$023$), NWO/OCW (Quantum Software Consortium No. $024$.$003$.$037$, Quantum Limits No. SUMMIT.$1$.$1016$), and from the Dutch Ministry of Economic Affairs (Quantum Delta NL).

\subsection*{Data availability}
The model underlying this manuscript and example simulations are available in Ref. \cite{Data}.

\bibliographystyle{unsrt}
\bibliography{alignmentbib} 

\end{document}